\documentstyle[12pt,graphicx,subfigure,amsmath]{article}
\def\lsim{\ ^<\llap{$_\sim$}\ }
\def\gsim{\ ^>\llap{$_\sim$}\ }
\def\r2{\sqrt 2}
\def\beq{\begin{equation}}
\def\eeq{\end{equation}}
\def\beqn{\begin{eqnarray}}
\def\eeqn{\end{eqnarray}}
\def\sinW2{\sin^2\theta_W}

\def\mz2{M_{z}^2}
\def\c2b{\cos 2\beta}

\def\fneut#1{{\tilde \chi}^{0}_#1}

\def\mx#1{m_{{\tilde \chi}^{0}_#1}}

\def\mz{M_Z}

\def\sec2w{sec^2\theta_W}
\def\gmin2{(g-2)_\mu}

\catcode`\@=11 % This allows one to modify PLAIN macros.

\def\lsim{\mathrel{\mathpalette\@versim<}}
\def\gsim{\mathrel{\mathpalette\@versim>}}
\def\@versim#1#2{\vcenter{\offinterlineskip
    \ialign{$\m@th#1\hfil##\hfil$\crcr#2\crcr\sim\crcr } }}

\begin{document}
\begin{flushright}
{TIFR/TH/03-10}
\end{flushright}
\begin{center}
  { \Large\bf Higgsino Dark Matter in a SUGRA Model with 
Nonuniversal Gaugino Masses 
\\}
  \vglue 0.5cm
  Utpal Chattopadhyay$^{(a)}$ and
    D.P. Roy $^{(b)}$
    \vglue 0.2cm
    {\em $^{(a)}$Department of Theoretical Physics, Indian Association 
for the Cultivation of Science, Raja S.C. Mullick Road, Kolkata 700032, India \\}
    {\em $^{(b)}$Department of Theoretical Physics, Tata Institute
      of Fundamental Research, Homi Bhabha Road, Mumbai 400005, India \\} 
  \end{center}        
\begin{abstract}
We study a specific SUGRA model with nonuniversal gaugino masses 
as an alternative to the minimal SUGRA model in the context of 
supersymmetric dark matter. 
The lightest supersymmetric particle 
in this model comes out to be a Higgsino dominated 
instead of a bino dominated lightest neutralino. 
The thermal relic density of this Higgsino dark matter is somewhat lower than 
the cosmologically favoured range, which means it may be only a subdominant 
component of the cold dark matter.  Nonetheless, it predicts favourable 
rates of indirect detection, which can be seen in square-km size neutrino 
telescopes.  
\end{abstract}
\section{Introduction}
 
The lightest supersymmetric particle (LSP)
in the standard R-parity conserving supersymmetric 
model is the leading particle physics candidate for the dark matter (DM) 
of the universe~\cite{Jungman:1995df}.  The most popular 
supersymmetry (SUSY) breaking 
model is the minimal supergravity (SUGRA) model having universal scalar, 
gaugino masses and trilinear couplings 
at the GUT scale.  Over most of the parameter space of this model the LSP 
is dominantly a bino ($\fneut1 \simeq {\tilde B} $) 
which does not couple to W or Z-boson.  Hence they can only pair-annihilate 
via the exchange of superparticles like squarks or sleptons, 
$\tilde B \tilde B \xrightarrow{\tilde q (\tilde l)} 
{ q \bar q (l^+  l^-) }$.  The experimental limits on these particle masses, 
$m_{\tilde q} \gsim 200 {\rm ~GeV}$ and 
$m_{\tilde l} \gsim 100 {\rm ~GeV}$~\cite{Hagiwara:fs}, 
imply a rather slow rate of pair annihilation. Consequently, 
the model predicts 
an over-abundance of the DM relic density over 
most of the parameter 
space~\cite{etcEllis:2001qm}.  This has led to several recent works, extending 
the SUSY DM investigations to nonminimal SUGRA 
models~[\ref{etcArnowitt:2001ca}-\ref{Bertin:2002sq}].  
%\cite{etcArnowitt:2001ca,Ellis:2002iu,
%Corsetti:2000yq,Bertin:2002sq}.
  While 
many of them explore models with nonuniversal scalar masses, we shall 
concentrate here on nonuniversal gaugino mass models.  In particular, we 
shall focus on a model leading generically to a Higgsino-like LSP. 
Because of its unsuppressed coupling to W and Z bosons the 
$\tilde H \tilde H \rightarrow W^+ W^-,ZZ $ annihilation rates via s-channel 
Z-boson and t-channel Higgsino exchanges are large. Besides, there is a near 
degeneracy of the lighter neutralinos and lightest chargino masses in 
this case, 
\beqn
\mx1 \simeq \mx2 \simeq m_{{\tilde \chi}_1^\pm} \simeq |\mu|, 
\label{degenerateeqn}
\eeqn
where $\mu$ is the supersymmetric Higgsino mass parameter.  This leads 
to large coannihilation cross sections~\cite{Mizuta:1992qp}.  Consequently, 
the Higgsino DM density falls below the cosmologically favoured 
range~\cite{etcBennett:2003bz}, 
\beqn
0.05 < \Omega_m h^2 < 0.2,
\label{cdmlim}
\eeqn
where the lower limit corresponds to the galactic density of 
DM ($\Omega_m \simeq 0.1 $) from rotation curves.  Thus the Higgsino 
DM can only be a subdominant component of the galactic DM density.  
However, its large coupling to Z implies a large rate of capture 
inside the Sun.  Hence, the model predicts a sizable indirect detection 
rate of Higgsino DM via high energy neutrinos coming from their 
pair annihilation in the solar core.  This is much larger than the 
minimal SUGRA model prediction and should be detectable at the future 
neutrino telescopes, as shown below.

\section{Non-Universal Gaugino Mass Model}
SUGRA model with nonuniversal gaugino masses at the GUT scale have been 
discussed in many earlier works~\cite{etcEllis:1985jn,
etcAnderson:1996bg,Huitu:1999vx}.  We shall only quote the main results 
here, focusing on the SU(5) GUT.  In this model the gauge kinetic 
function depends on a nonsinglet chiral superfield $\Phi$, whose 
auxiliary $F$-component acquires a large vacuum expectation value (vev). 
Then the gaugino masses 
come from the following dimension five term in the Lagrangian: 
\beqn
L={\frac{{<F_\Phi>}_{ij}}{M_{Planck}}} \lambda_i \lambda_j
\eeqn  
where $\lambda_{1,2,3}$ are the $U(1)$, $SU(2)$ and $SU(3)$ gaugino fields 
i.e. the bino $\tilde B$, the wino $\tilde W$ and the gluino 
$\tilde g$ respectively.
Since the gauginos belong to the adjoint representation of SU(5), $\Phi$ and 
$F_\Phi$ can belong to any of the irreducible representations appearing 
in their symmetric product, i.e. 
\beqn
{(24 \times 24)}_{symm} =1+24+75+200
\eeqn
The minimal SUGRA model assumes $\Phi$ to be a singlet, which 
implies equal gaugino masses at the GUT scale. On the other hand if 
$\Phi$ belongs to one of the
nonsinglet representations of $SU(5)$, then these gaugino masses are unequal 
but related to one another via the representation invariants.  Thus the 
three gaugino masses at the GUT scale in a given representation $n$ 
are determined in terms of a single SUSY breaking mass parameter 
$M_{1/2}$ by 
\beqn
M^{G,n}_{1,2,3} = C^n_{1,2,3} M_{1/2}
\label{relativegauginos}
\eeqn
where $C^{1}_{1,2,3}=(1,1,1)$, $C^{24}_{1,2,3}=(-1,-3,2)$, $C^{75}_{1,2,3}=(-5,3,1)$ and
$C^{200}_{1,2,3}=(10,2,1)$.  The resulting ratios of 
$M_i^G$'s for each $n$ are listed 
in Table~\ref{tabrelativeweights}.
\begin{table}[h]
%\vspace{3mm}
{\centering
\begin{tabular}[h]{|c| c c c|}
\hline
$n$ & $M^G_3$ & $M^G_2$ & $M^G_1$ \\
\hline
\hline
1 & 1 & 1 & 1 \\
24 & 1 & $-3/2$ & $-1/2$ \\
75 & 1 & 3 & $-5$  \\
200 & 1 & 2 & 10  \\
\hline 
\end{tabular}
\par}
\centering
\caption{Relative values of the SU(3), SU(2) and U(1) 
gaugino masses at GUT scale for different 
representations $n$ of the chiral superfield $\Phi$.}
\label{tabrelativeweights}
\end{table}

\noindent
Of course in general the gauge kinetic function can involve several chiral 
superfields belonging to different representations of $SU(5)$ 
which gives us the freedom to vary mass ratios continuously.  We shall explore 
such a possibility in a future work.  But let us concentrate here on the 
representations 1, 24, 75 and 200 individually.  While the singlet 
representation corresponds to universal gaugino masses, each of the 
nonsinglet representations corresponds to definite mass ratios and is 
therefore as predictive as the former. 

  	These nonuniversal gaugino mass models are known to be consistent 
with the observed universality of the gauge couplings at the GUT 
scale~[\ref{etcEllis:1985jn}--\ref{Chattopadhyay:2001mj}]
%~\cite{etcEllis:1985jn,etcAnderson:1996bg,
%Huitu:1999vx,Chattopadhyay:2001mj}, 
\beqn
\alpha_3^G=\alpha_2^G=\alpha_1^G=\alpha^G (\simeq 1/25)
\eeqn  
Since the gaugino masses evolve like the gauge couplings at one loop level 
of the renormalisation group equations (RGE), the 
three gaugino masses at the electroweak (EW) 
scale are proportional to the corresponding gauge couplings, i.e.
\beqn
M_1 & = & (\alpha_1/\alpha_G) M_1^G  \simeq  (25/60) C_1^n M_{1/2} \nonumber \\
M_2 & = & (\alpha_2/\alpha_G) M_2^G \simeq  (25/30)C_2^n M_{1/2} \nonumber \\ 
M_3 & = & (\alpha_3/\alpha_G) M_3^G  \simeq  (25/9) C_3^n M_{1/2}  
\label{gauginoEW}
\eeqn
 
For simplicity we shall assume a universal SUSY breaking scalar mass $m_0$ 
at the GUT scale.  Then the corresponding scalar masses at the EW scale are 
given by the renormalisation group evolution formulae~\cite{Carena:1994bv}.
  A very important SUSY breaking mass parameter at this scale is 
$m_{H_2}$ , as it appears in the EW symmetry breaking condition, 
\beqn
\mu^2 + M_Z^2/2= {{m_{H_1}^2 -m_{H_2}^2\tan^2 \beta} \over {\tan^2\beta -1} }
\simeq -m_{H_2}^2,
\eeqn
where the last equality holds for the $\tan\beta \gsim$ 5 region, which is 
favoured by the Higgs mass limit from LEP~\cite{Hagiwara:fs}.  Expressing 
$m_{H_2}^2$ 
at the right hand side in terms of the GUT scale mass parameters at a 
representative value of $\tan\beta=10$ 
gives~\cite{Carena:1994bv,Komine:2000tj}
\beqn
\mu^2 + \frac{1}{2} M_Z^2  & \simeq &  
-0.1m_0^2 +2.1 {M_3^G}^2 -0.22 {M_2^G}^2 
-0.006{M_1^G}^2 +0.006 M_1^G M_2^G +  \nonumber \\ 
& & 0.19 M_2^G M_3^G + 0.03 M_1^G M_3^G,
\label{musqeqn}
\eeqn
neglecting the contribution from the trilinear coupling term at the 
GUT scale.  Moreover, the coefficients vary rather mildly over the
 moderate $\tan\beta$ region.  Although we shall use exact numerical 
solutions to the two-loop RGE, two points are worth noting from this simple 
equation. 

Firstly, eq.\eqref{musqeqn} gives a measure of fine-tuning from the 
required degree of cancellation between the dominant terms $\mu^2$ and 
${M_3^G}^2$ to give the right EW scale $M_Z^2/2$.  The LEP  
limit on the lightest chargino mass, $m_{{\tilde \chi}_1^\pm} > 100
{~\rm GeV}$~\cite{Hagiwara:fs} implies 

\beqn
|\mu|,|M_2| > 100~{\rm GeV} 
\label{charginolimit}
\eeqn
while eq.\eqref{musqeqn} implies
\beqn
\mu^2 + M_Z^2/2 \simeq 2.1 M_{1/2}^2 ~({\rm ~for ~mSUGRA}) {\rm ~and,} \nonumber \\
\mu^2 + M_Z^2/2 \simeq 1.4 M_{1/2}^2 ~({\rm ~for ~n=200 ~model}) 
\label{mufromm2}
\eeqn
Thus for the universal gaugino mass case of mSUGRA  
eqs.(\ref{gauginoEW}), (\ref{charginolimit}) and (\ref{mufromm2}) 
imply fine-tuning at least at the level of 
$\simeq 10$~\cite{Barbieri:1987fn}.  On 
the other hand one sees from these equations that the 
fine-tuning problem is significantly alleviated in the nonuniversal models 
with $n=75$ and 200, corresponding to $C^n_2=3$ and 2 respectively~\cite{Kane:1998im}. 

    Secondly, the universal gaugino mass model corresponds to 
$ |\mu| > |M_2| > |M_1| $, which implies that the lighter chargino and 
neutralinos are dominantly gauginos with hierarchical masses -i.e. 
$m_{\chi_1^0} \simeq M_1$ and $m_{\chi_2^0,\chi_1^\pm} \simeq M_2\simeq 2 M_1$.  
There is however a narrow strip of very large $m_0$ region where the 
first term of eq.(\ref{musqeqn}) pushes down $|\mu| $ towards the LEP 
limit as given in eq.(\ref{charginolimit}).  
This is the so called focus point region~\cite{etcFeng:1999zg}, where the 
lighter chargino and neutralinos are mixed Higgsino-gaugino states.  But, 
over the bulk of the parameter space the LSP is dominantly a 
$\tilde B$, which leads to an over-abundance of the DM relic density, 
as discussed earlier.  One expects from Table~1 a similar result 
for the $n=24$ model.  In fact it predicts a larger hierarchy 
between the $\chi_1^0$ and $\chi_2^0 (\chi_1^\pm)$ masses, $M_1$ and 
$M_2$. Consequently the LSP is completely dominated by $\tilde B$ and can be 
relatively light.  The SUSY DM phenomenology for this case has been discussed 
recently in Ref.~\cite{Corsetti:2000yq,Bertin:2002sq,Belanger:2001am}.  
In contrast we see from eqs.(\ref{gauginoEW}), (\ref{mufromm2}) and 
Table~1 that the $n=200$ model predict the opposite 
hierarchy $|\mu| < |M_2| < |M_1|$, while the $n=75$ model has 
$|\mu| < |M_1| < |M_2|$.  Thus the lighter chargino and 
neutralino states are Higgsino dominated and roughly degenerate 
(eq.(\ref{degenerateeqn})) over the bulk of the parameter space for 
both $n=75$ and 200 models.  This leads to a DM relic density 
somewhat below the cosmologically favoured range (eq.(\ref{cdmlim})). 
We should mention here that unlike the case of mSUGRA, there is no 
possibility of having stau coannihilations in 
the $n=75$ and the $n=200$ models.  This is related to 
staus being significantly heavier than $m_{\tilde \chi_1^0}$ 
in these scenarios for all values of $\tan\beta$. This in fact  
originates from the gauge sector running of the slepton RGEs 
due to the specific gaugino mass nonuniversalities. 

  It is for the above reason that the Higgsino DM phenomenology of the 
nonuniversal SUGRA models, corresponding to $n=75$ and 200 have not been 
explored in detail so far.  We feel that this is important for two 
reasons: (i) Even though the Higgsino may be a subdominant component 
of the galactic DM density, its large coupling to  $Z$-boson 
implies large rate of capture inside the Sun.  Hence the model predicts a 
sizable indirect detection rate via high energy neutrinos coming from 
their pair annihilation inside the sun.  Even after rescaling by the low 
DM density factor the indirect detection rate comes out to be larger than the 
minimal SUGRA predictions~\cite{Barger:2001ur}.  
(ii)  It is possible that the thermal
 relic density of the Higgsino DM is enhanced by either a modification of
 the freeze-out temperature of the standard cosmological model 
due to a quintessence field as
suggested in Ref.\cite{etcSalati:2002md} or by nonthermal 
production mechanisms of the type suggested in 
Ref.\cite{etcMurakami:2000me}.  In that case it can be the 
dominant component of the galactic DM density.  Therefore we have computed
the indirect and direct detection rates both with and without 
rescaling.  
   
\section{Higgsino Dark Matter in the n=200 SUGRA Model}
\begin{figure}           
\vspace*{-1.0in}                                 
%\subfigure[Caption for picture one]{                       
\subfigure[]{                       
\label{relic_a} 
\hspace*{-0.6in}                     
\begin{minipage}[b]{0.5\textwidth}                       
\centering
\includegraphics[width=\textwidth,height=0.8\textwidth]
{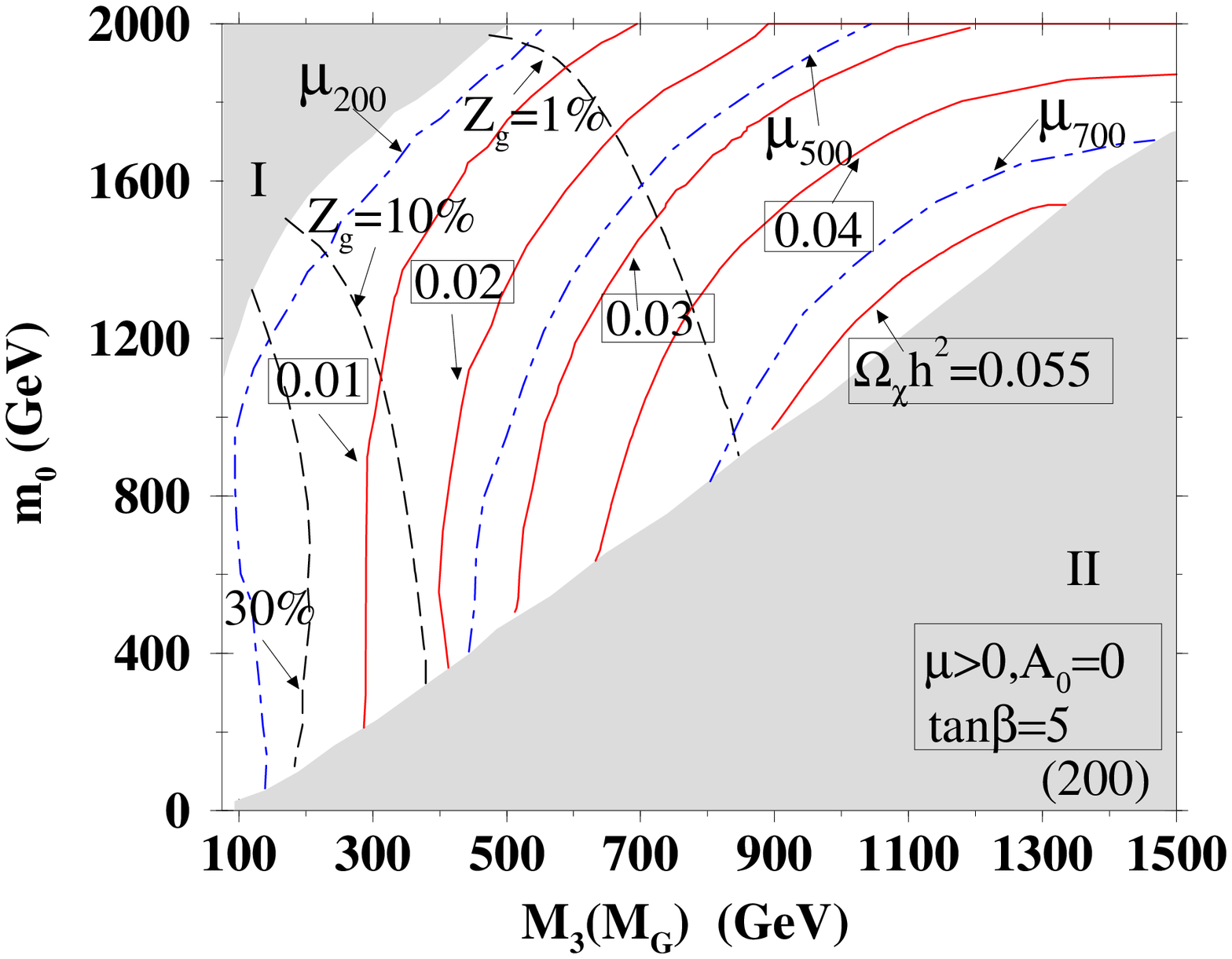}    
\end{minipage}}                       
\hspace*{0.3in}
%\subfigure[Caption for picture two]{                       
\subfigure[]{                       
\label{relic_b}                       
\begin{minipage}[b]{0.5\textwidth}                       
\centering                      
\includegraphics[width=\textwidth,height=0.8\textwidth]
{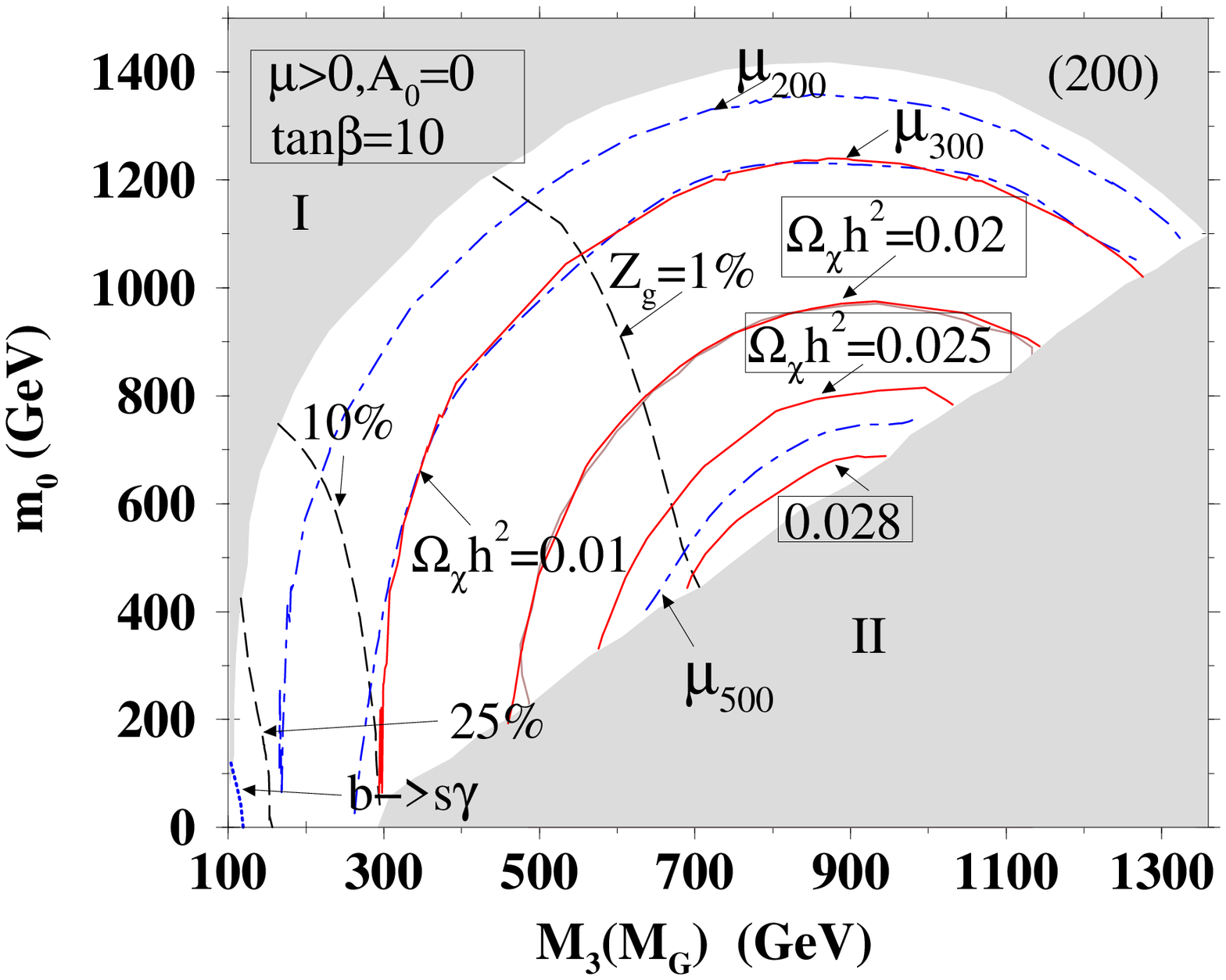}    
\end{minipage}}                       
%\caption{Caption for both of the pictures}                       
%\label{Label_for_both_of_the_figures1} 
\hspace*{-0.6in}                     
%\subfigure[Caption for picture one]{                       
\subfigure[]{                       
\label{relic_c}                      
\begin{minipage}[b]{0.5\textwidth}                       
\centering
\includegraphics[width=\textwidth,height=0.8\textwidth]
{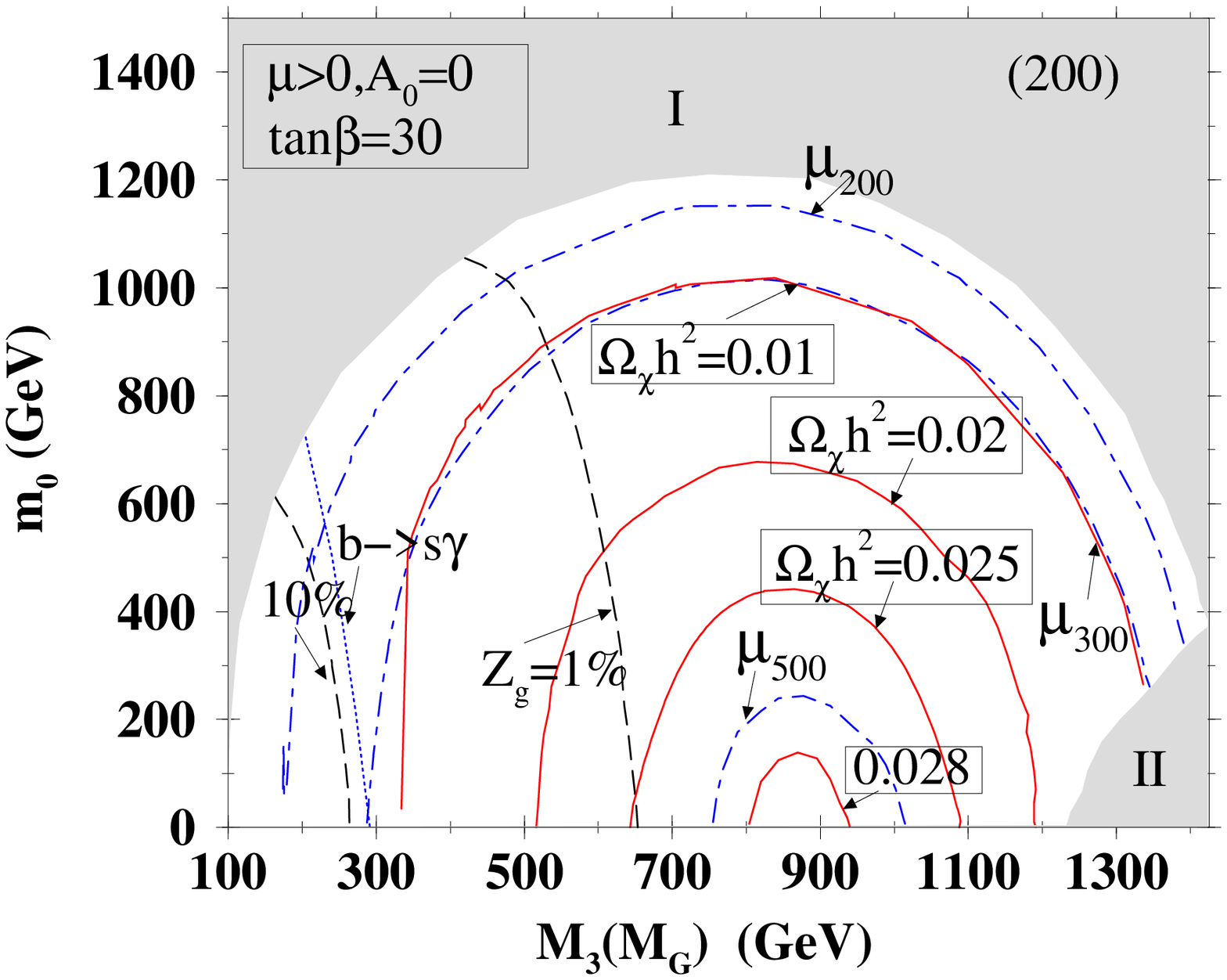}
\end{minipage}}
\hspace*{0.3in}                       
%\subfigure[Caption for picture two]{                       
\subfigure[]{                       
\label{relic_d}                       
\begin{minipage}[b]{0.5\textwidth}                       
\centering                      
\includegraphics[width=\textwidth,height=0.8\textwidth]
{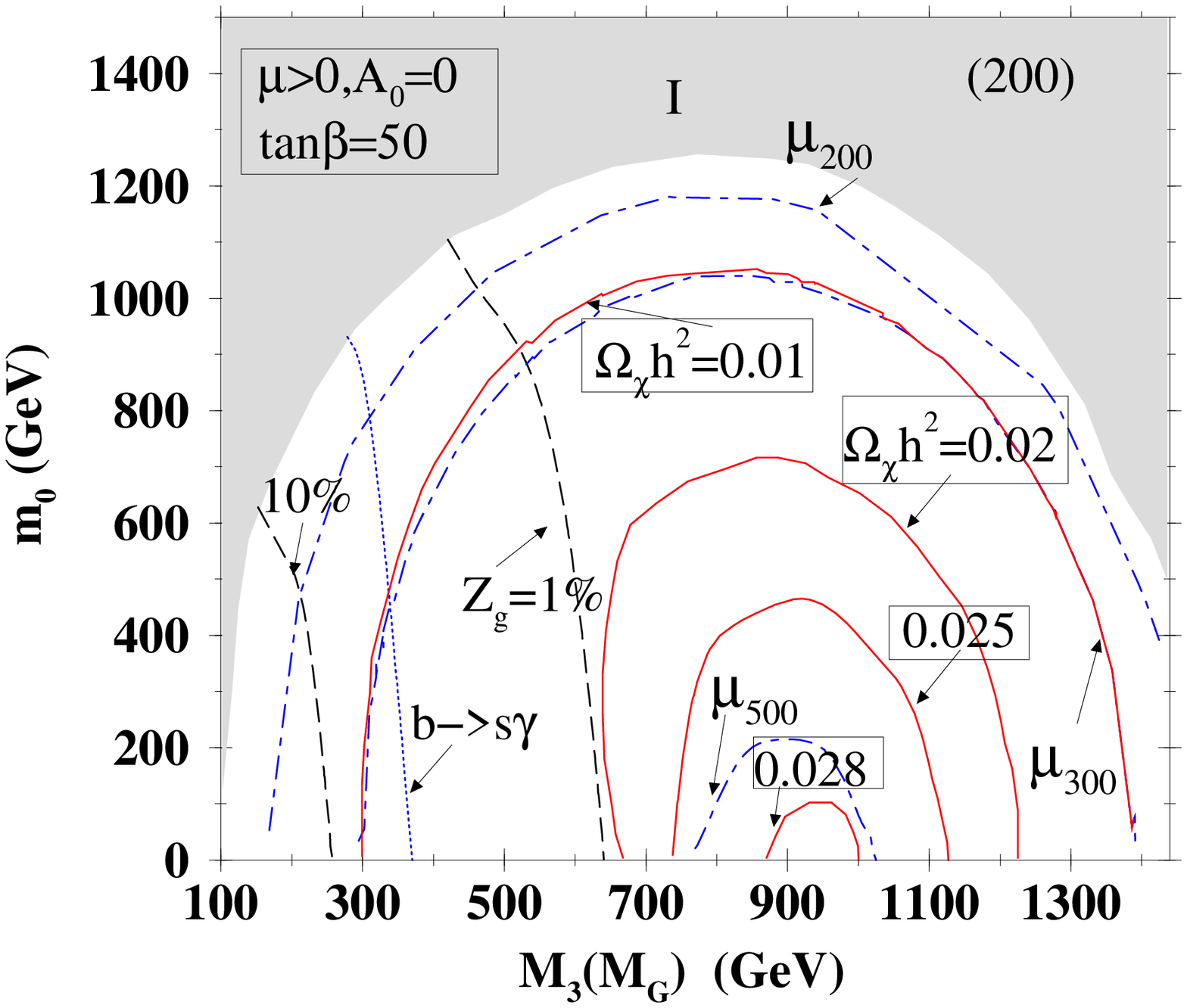}    
\end{minipage}}                       
\caption{Contours of $\Omega_{\tilde \chi} h^2$, 
$\mu$, $Z_g$ and $Br(b \rightarrow 
s+\gamma)$ in the ($m_0-M_{3}(M_G)$) 
plane for $\tan\beta=$ $5,10,30$ and $50$ for the 
$n=200$ nonuniversal gaugino mass model with $\mu>0$. 
The neutralino relic density $\Omega_{\tilde \chi} h^2$ contours are shown 
as (red) solid 
lines. The contours for $\mu$ are shown as (blue) dot-dashed lines where the 
notation $\mu_{500}$ means $\mu=500$ GeV. The gaugino fraction for 
the LSP  $Z_g$ contours are 
shown as (black) dashed lines. The $Br(b \rightarrow s+\gamma)$ 
excluded regions for each figure are the regions left to the 
(blue) dotted lines. The (gray) I-zones in the top left parts 
are discarded via the lighter chargino mass lower bound and 
the absence of radiative EWSB and (gray) II-zones are eliminated because 
of the lack of stability of the Higgs potential at the GUT scale $M_G$.
}                       
\label{relic_figs}

\end{figure}

We shall concentrate on the nonuniversal SUGRA model corresponding 
to $n=200$ because it can generate radiative electroweak symmetry breaking 
(EWSB) over a much wider range of parameters compared to the $n=75$ 
case, the latter being restricted to have  
small $\tan\beta$ solutions only.  
Fig.~\ref{relic_figs} shows the allowed regions in the 
$m_0-M_{1/2}(=M_3^G)$ plane for $\tan\beta=$5, 10, 30 and 50.  The area 
marked I at the top is disallowed because $\mu^2$ falls below 
the LEP limit (eq.\ref{charginolimit}) and then becomes negative, signalling 
the absence of EWSB.  The area marked II at the bottom is disallowed 
because the Higgs potential becomes unstable at the GUT scale.  We have 
chosen $\mu>0$ since the $\mu<0$ branch is strongly disfavoured by the 
$b \rightarrow s+ \gamma$ branching ratio, along with the muon anomalous 
magnetic moment ($a_\mu$) constraint.  This figure shows that the 
$b \rightarrow s+ \gamma$ constraint to be rather mild for $\mu>0$.  The
lower limit from $a_\mu$ (not shown) is even milder.  

We see from the contours of $\mu$ in Fig.~\ref{relic_figs} that one 
generally has $\mu <500$~GeV in the model as anticipated.  It also shows the
gaugino component of the LSP,
\beqn
Z_g=N_{11}^2 + N_{12}^2,
\label{gauginofrac}
\eeqn
where 
\beqn
{\tilde \chi}_1^0
=N_{11}\tilde B +N_{12}\tilde W + N_{13}{\tilde H}_1 + N_{14}{\tilde H}_2.
\label{neutalinocomposition}
\eeqn 
We see that the bulk of the allowed parameter space corresponds to 
$Z_g < 10 \%$, which means that the LSP is dominated by the Higgsino 
component to more that $90\%$.  Finally Fig.~\ref{relic_figs} shows the 
contours of neutralino relic density $\Omega_{\chi} h^2$  
which was computed using the micrOMEGAs of 
Ref.~\cite{Belanger:2001fz}.  It is seen to generally lie below 
the lower limit of the cosmologically desirable range of 
eq.(\ref{cdmlim}) by a factor of 2 to 4.  This is due to the rapid 
pair annihilation processes $\tilde H \tilde H \rightarrow W^+ W^-, 
ZZ$ via s-channel Z-boson and t-channel Higgsino exchanges, as mentioned 
earlier.  Besides, the near degeneracy of the lighter chargino and 
neutralino masses (eq.\ref{degenerateeqn}) leads to large coannihilations. 
  In view of the mass degeneracy it is important to include radiative 
corrections to the ${\tilde \chi}_{1,2}^0$ and ${\tilde \chi}_1^\pm$ masses 
in the Higgsino LSP scenario~\cite{Drees:1996pk}.  We have included 
this using the code of Manuel Drees. But, it does not enhance the 
neutralino relic density significantly.  It should be noted here that the 
dominant gaugino component of the LSP (eq.\ref{neutalinocomposition}) comes 
from $\tilde W$ instead of $\tilde B$, in view of the inverted mass 
hierarchy $\mu <M_2 <M_1$ in this model.  Since, the winos have very 
similar annihilation mechanisms like the Higgsinos there is no increase 
in the relic density in the mixed Higgsino-gaugino region ($Z_g > 10 \%$). 
\section{Indirect Detection Rates}   
\begin{figure}           
\vspace*{-1.0in}                                 
%\subfigure[Caption for picture one]{                       
\subfigure[]{                       
\label{phimu_a} 
\hspace*{-0.6in}                     
\begin{minipage}[b]{0.5\textwidth}                       
\centering
\includegraphics[width=\textwidth,height=0.8\textwidth]{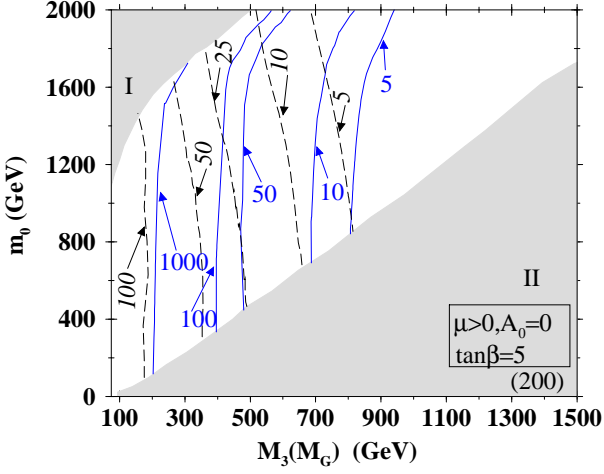}    
\end{minipage}}                       
\hspace*{0.3in}
%\subfigure[Caption for picture two]{                       
\subfigure[]{                       
\label{phimu_b}                       
\begin{minipage}[b]{0.5\textwidth}                       
\centering                      
\includegraphics[width=\textwidth,height=0.8\textwidth]
{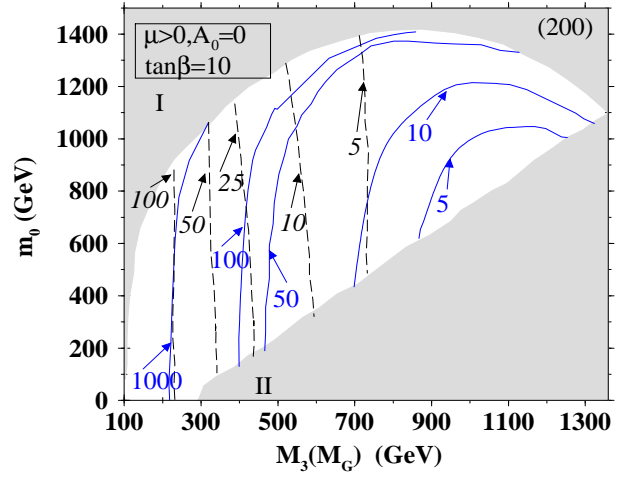}    
\end{minipage}}                       
%\caption{Caption for both of the pictures}                       
%\label{Label_for_both_of_the_figures1} 
\hspace*{-0.6in}                     
%\subfigure[Caption for picture one]{                       
\subfigure[]{                       
\label{phimu_c}                      
\begin{minipage}[b]{0.5\textwidth}                       
\centering
\includegraphics[width=\textwidth,height=0.8\textwidth]{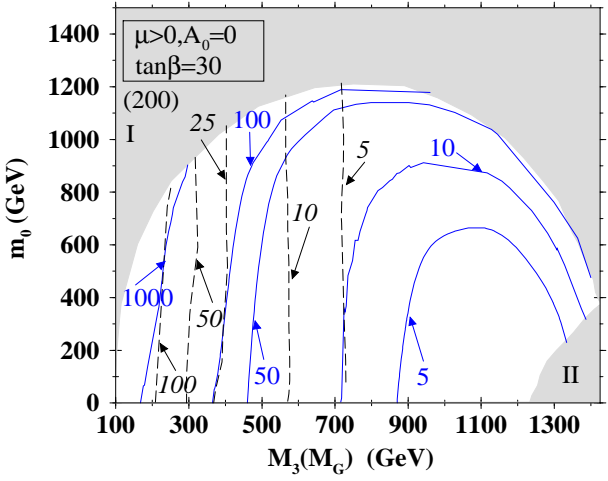}
\end{minipage}}
\hspace*{0.3in}                       
%\subfigure[Caption for picture two]{                       
\subfigure[]{                       
\label{phimu_d}                       
\begin{minipage}[b]{0.5\textwidth}                       
\centering                      
\includegraphics[width=\textwidth,height=0.8\textwidth]{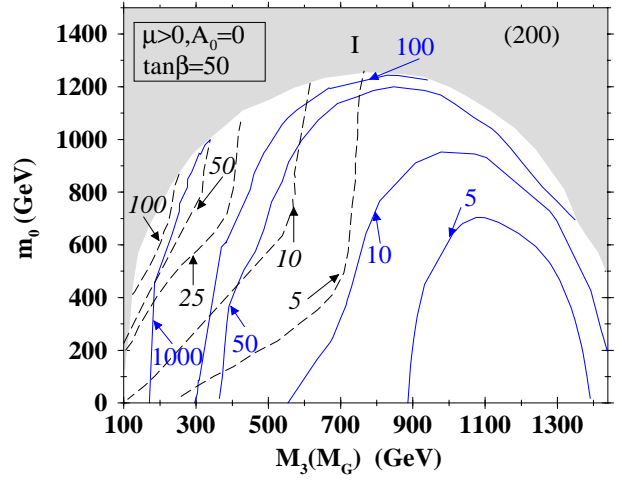}    
\end{minipage}}                       
\caption{
Contours for muon flux from the Sun in units of 
${\rm km}^{-2} {\rm yr}^{-1}$ for $\tan\beta=$ 5, 10, 30 and 50 in the 
nonuniversal gaugino mass model with $n=200$.  
The (blue) solid lines refer to $\phi_\mu$ and 
the (black) dashed lines correspond to $\xi \phi_\mu$ where $\xi=
{\Omega_{\tilde \chi} h^2}/
{<\Omega_{\tilde \chi} h^2>}_{min}$, with 
${<\Omega_{\tilde \chi} h^2>}_{min}=0.05$.  
The regions marked with I and II are 
same as that in Fig.{\ref{relic_figs}}       
}                       
\label{phimu} 

\end{figure}

Since the Z-boson couples only to the Higgsino component of neutralino, the
$Z {\tilde \chi}_1^0 {\tilde \chi}_1^0 $ coupling is proportional to 
$N_{13,14}^2$~\cite{Guchait:1994zk}.  Moreover, the spin dependent force 
from Z-exchange is known to dominate the ${\tilde \chi}_1^0$ interaction 
rate with the solar matter, which is predominantly 
Hydrogen~\cite{Jungman:1995df}.  Hence the solar capture rate of the 
Higgsino DM is predicted to be enormously larger than the bino DM of the 
minimal SUGRA model.  This implies in turn an enormously higher rate of pair
annihilation, $\tilde H \tilde H \rightarrow W^+ W^- (ZZ)$, since the 
capture and annihilation rates balance one another at equilibrium.  The 
high energy neutrinos coming from W(Z) decay are expected to be detected at 
the large area neutrino telescopes like the IceCubes~\cite{Ahrens:2002dv} and
the ANTARES~\cite{Aslanides:1999vq} via their charged current interaction 
($\nu \rightarrow \mu$).  The resulting muons constitute the so called 
indirect DM detection signal.  Fig.~\ref{phimu} shows the indirect signal 
rate contours over the full parameter space which was computed by using 
DARKSUSY of Ref.~\cite{Gondolo:2002tz}.  The signal contours are shown 
both with and without rescaling by a factor 
$\xi=\Omega_\chi h^2/0.05$~\cite{Bottino:2001it}.  The denominator 
corresponds to $\Omega_m=0.1$ which is the galactic DM 
relic density assumed in this computation.  Even with rescaling one expects 
muon flux $\phi_\mu$ of 5-100 ${\rm ~events/km^2/year}$ over practically 
the full parameter space of the model.  In contrast the minimal SUGRA     
model predicts a $\phi_\mu <1 {\rm ~event/km^2/year} $ over the 
entire parameter space, except for a very narrow strip at the 
$|\mu|=100 {\rm ~GeV}$ boundary corresponding to a larger 
Higgsino content in the LSP~\cite{Barger:2001ur}.  The proposed 
large area neutrino telescopes like IceCube and ANTARES are expected 
to cover a detection area of 1${\rm ~km}^2$.  The irreducible background 
for these experiments, coming from the high energy neutrinos produced 
by the cosmic ray interaction with the solar corona is estimated to be 
$\phi_\mu \sim 5 {\rm ~events/km^2/year}$~\cite{Barger:2001ur}.  Therefore
these experiments can probe a signal of $\phi_\mu \gsim 5 
{\rm ~events/km^2/year}$, as expected over practically the 
full parameter space of this nonuniversal SUGRA model.  It may be 
mentioned here that in the presence of some enhancement mechanism for 
the Higgsino DM relic 
density~\cite{etcSalati:2002md,etcMurakami:2000me}, it can become 
the dominant component of the cold dark matter.  This will enhance the 
indirect detection rate further, as indicated by the contours without 
rescaling. 
\section{Direct Detection Rates}     
For the sake of completeness we have computed the ${\tilde \chi^0_1}p$ elastic 
scattering cross-sections in this model, which determine the signal rate 
in direct detection experiments.  Both the spin-dependent and the 
spin-independent cross-sections have been computed using the DARKSUSY 
code~\cite{Gondolo:2002tz}. 

\begin{figure}           
\vspace*{-1.0in}                                 
%\subfigure[Caption for picture one]{                       
\subfigure[]{                       
\label{sigmasd_a} 
\hspace*{-0.6in}                     
\begin{minipage}[b]{0.5\textwidth}                       
\centering
\includegraphics[width=\textwidth,height=0.8\textwidth]
{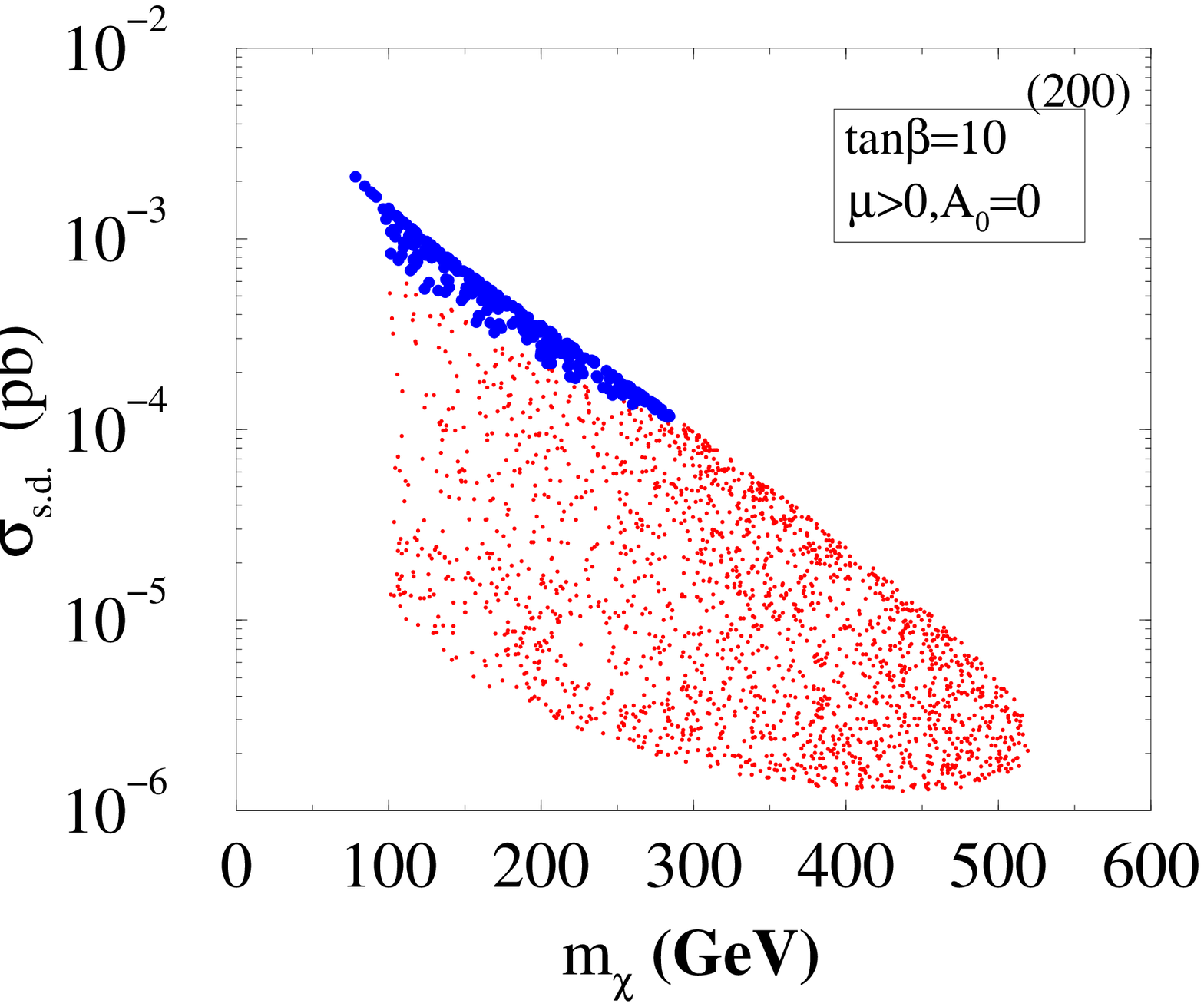}    
\end{minipage}}                       
\hspace*{0.3in}
%\subfigure[Caption for picture two]{                       
\subfigure[]{                       
\label{sigmasd_b}                       
\begin{minipage}[b]{0.5\textwidth}                       
\centering                      
\includegraphics[width=\textwidth,height=0.8\textwidth]
{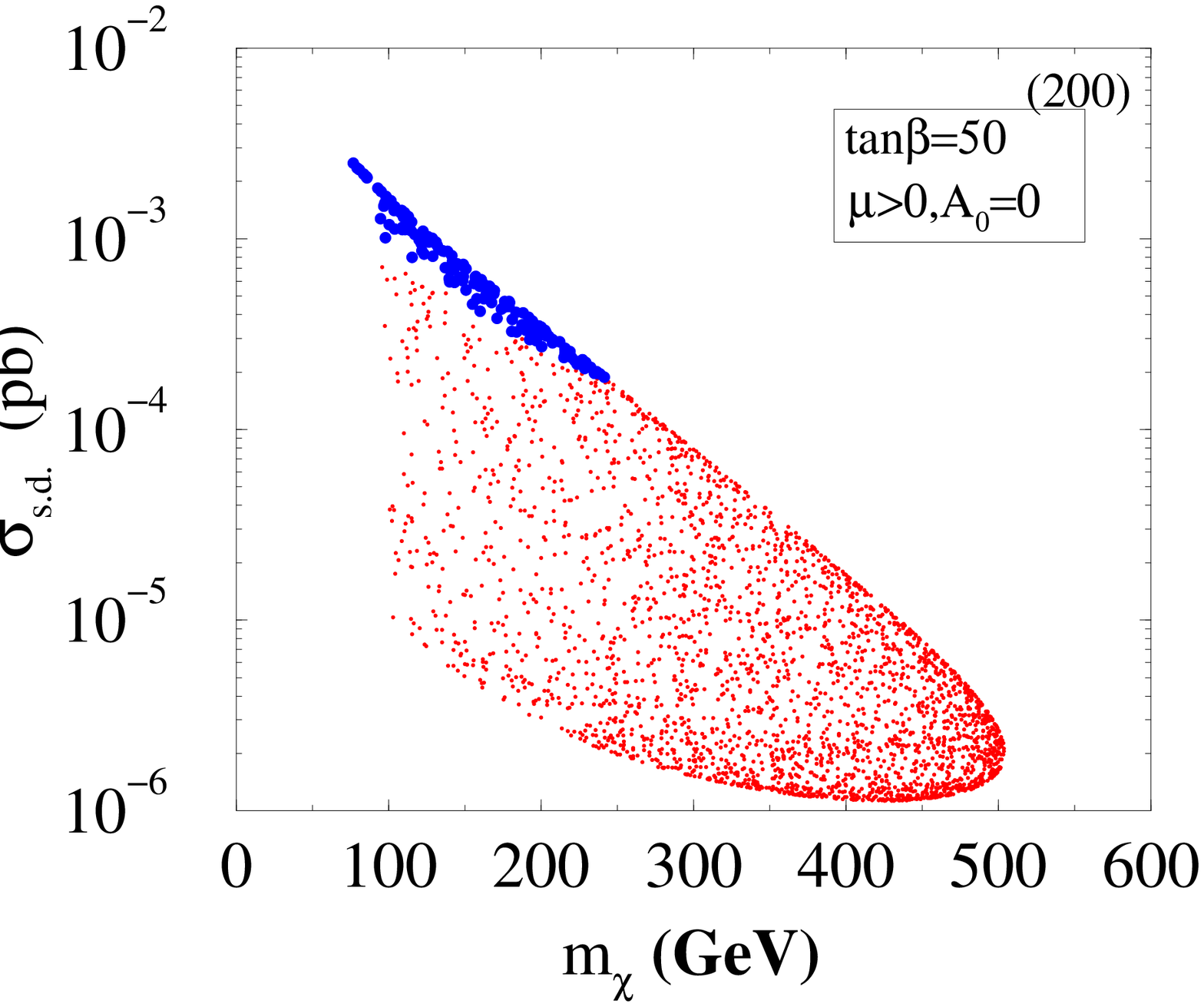}    
\end{minipage}}                       
%\caption{Caption for both of the pictures}                       
%\label{Label_for_both_of_the_figures1} 
\hspace*{-0.6in}                     
%\subfigure[Caption for picture one]{                       
\subfigure[]{                       
\label{sigmasd_c}                      
\begin{minipage}[b]{0.5\textwidth}                       
\centering
\includegraphics[width=\textwidth,height=0.8\textwidth]{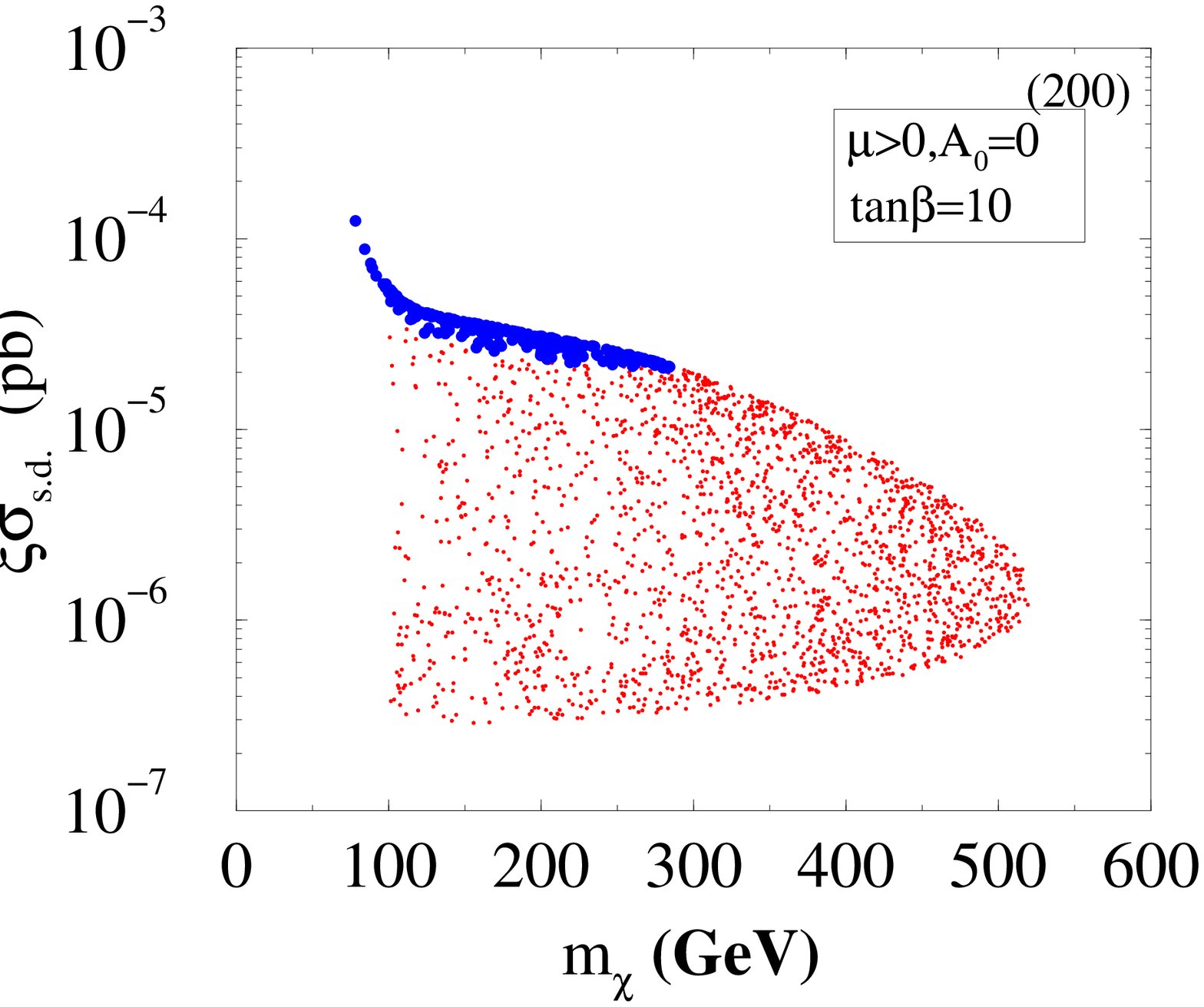}
\end{minipage}}
\hspace*{0.3in}                       
%\subfigure[Caption for picture two]{                       
\subfigure[]{                       
\label{sigmasd_d}                       
\begin{minipage}[b]{0.5\textwidth}                       
\centering                      
\includegraphics[width=\textwidth,height=0.8\textwidth]{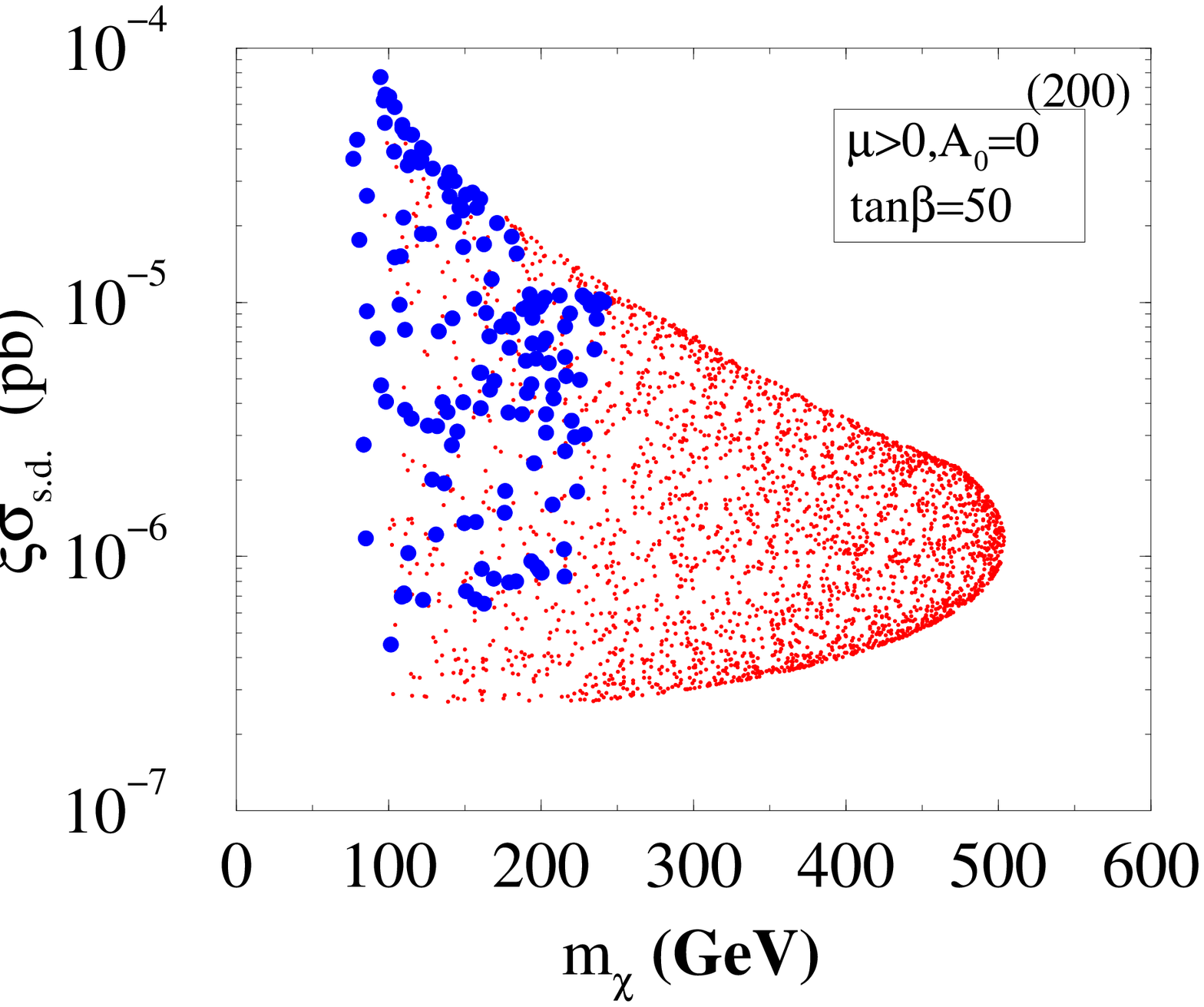}    
\end{minipage}}                       
\caption{(a) and (b): Spin dependent neutralino-proton 
scattering cross sections $\sigma_{{\tilde {\chi_1^0}}-p}^{SD}$  
vs $m_{{\chi_1}^0}$ for 
$\tan\beta=$ 10 and 50 respectively for the 
$n=200$ nonuniversal gaugino mass model.  The (red) dots and the (blue) filled 
circles refer to points with below and above 10\% gaugino 
fractions respectively in this generic Higgsino dominated 
LSP scenario.
(c) and (d): Same as (a) and (b) except for the scaled cross sections 
$\xi \sigma_{ {\tilde \chi}_1^0 -p }^{SD}$, where $\xi= 
{\Omega_{\tilde \chi} h^2}/
{<\Omega_{\tilde \chi} h^2>}_{min}$, with 
${<\Omega_{\tilde \chi} h^2>}_{min}=0.05$.}                          
\label{sigmasd} 

\end{figure}

The spin-dependent cross section is known to be dominated by Z-exchange. 
Therefore the spin-dependent cross-sections are much larger here compared 
to the minimal SUGRA model.  Fig.~\ref{sigmasd} gives scatter plots of 
spin-dependent cross section against the LSP mass, both with and without 
rescaling for $\tan\beta=$10 and 50.  Even the rescaled cross-section 
is 1-2 orders of magnitude larger than the minimal SUGRA 
predictions~\cite{Ellis:2001pa}.  Unfortunately, the direct detection 
experiments are not sensitive to the spin-dependent cross-section as they 
are based on heavy nuclei.  For example the UKDMC detector is only 
sensitive to a spin-dependent cross section 
$\gsim 0.5 {\rm pb}$~\cite{Ellis:2001pa}, which is much above any SUSY 
model prediction. 

\begin{figure}           
\vspace*{-1.0in}                                 
%\subfigure[Caption for picture one]{                       
\subfigure[]{                       
\label{sigma_a} 
\hspace*{-0.6in}                     
\begin{minipage}[b]{0.5\textwidth}                       
\centering
\includegraphics[width=\textwidth,height=0.8\textwidth]
{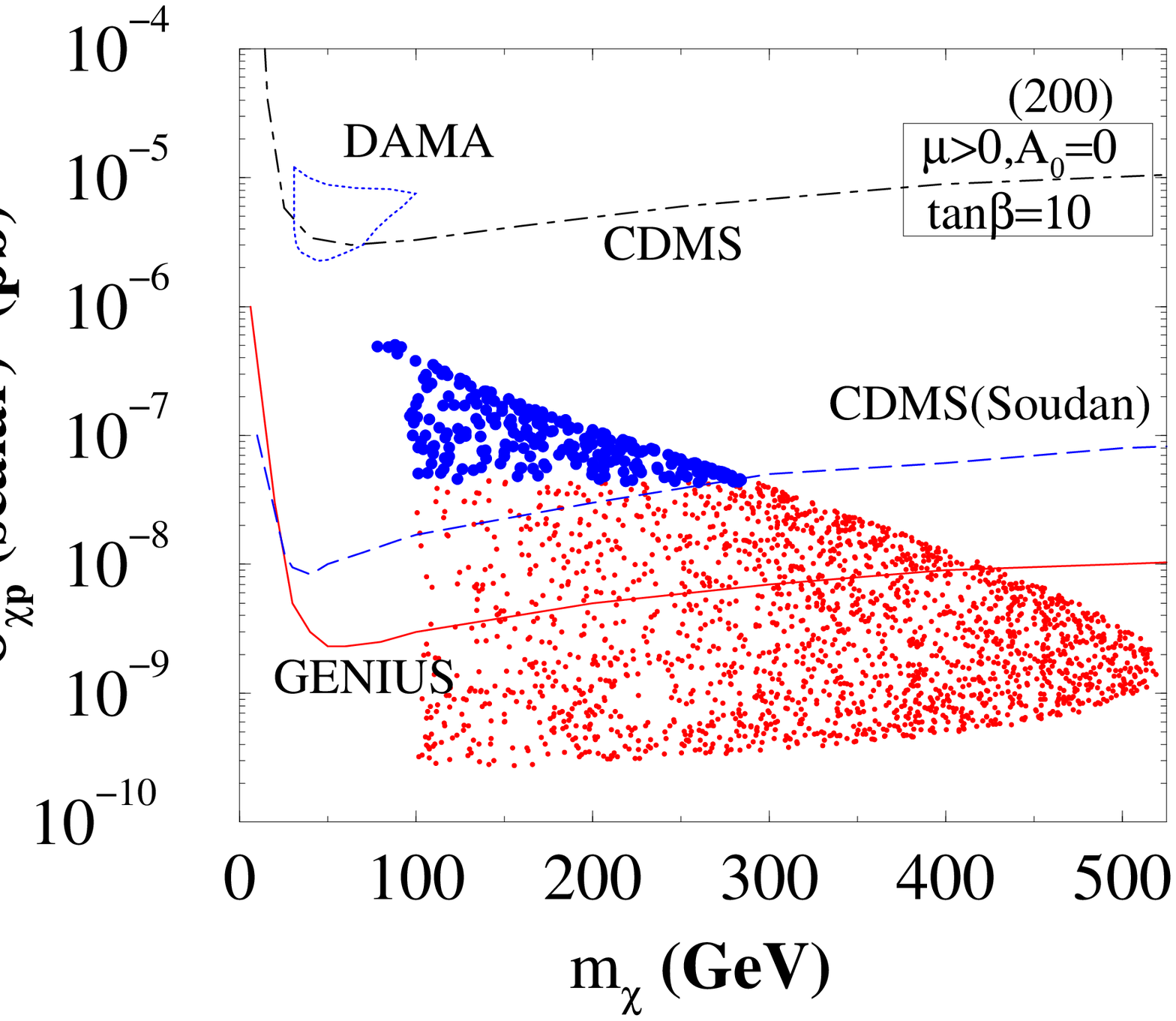}    
\end{minipage}}                       
\hspace*{0.3in}
%\subfigure[Caption for picture two]{                       
\subfigure[]{                       
\label{sigma_b}                       
\begin{minipage}[b]{0.5\textwidth}                       
\centering                      
\includegraphics[width=\textwidth,height=0.8\textwidth]
{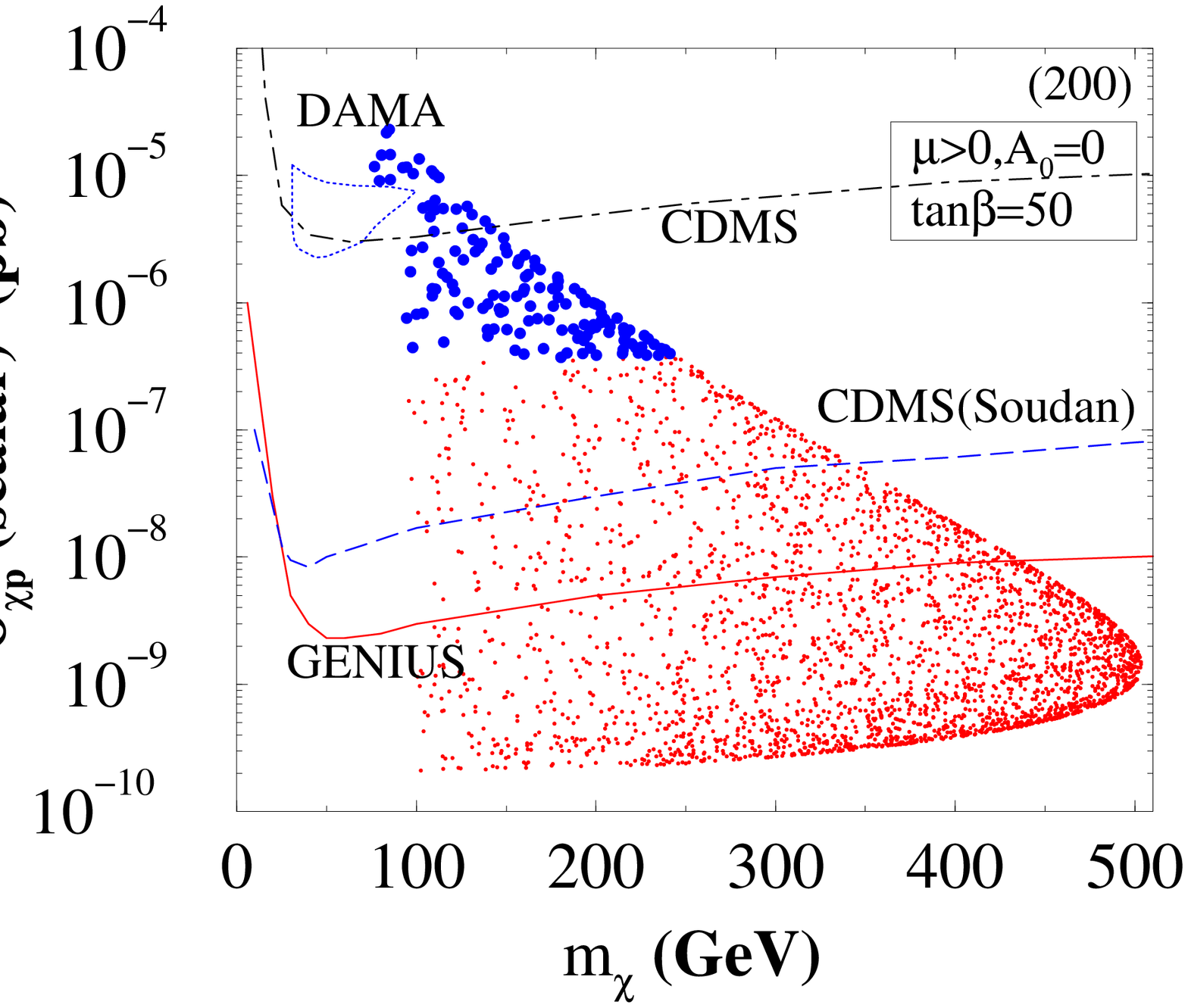}    
\end{minipage}}                       
%\caption{Caption for both of the pictures}                       
%\label{Label_for_both_of_the_figures1} 
\hspace*{-0.6in}                     
%\subfigure[Caption for picture one]{                       
\subfigure[]{                       
\label{sigma_c}                      
\begin{minipage}[b]{0.5\textwidth}                       
\centering
\includegraphics[width=\textwidth,height=0.8\textwidth]{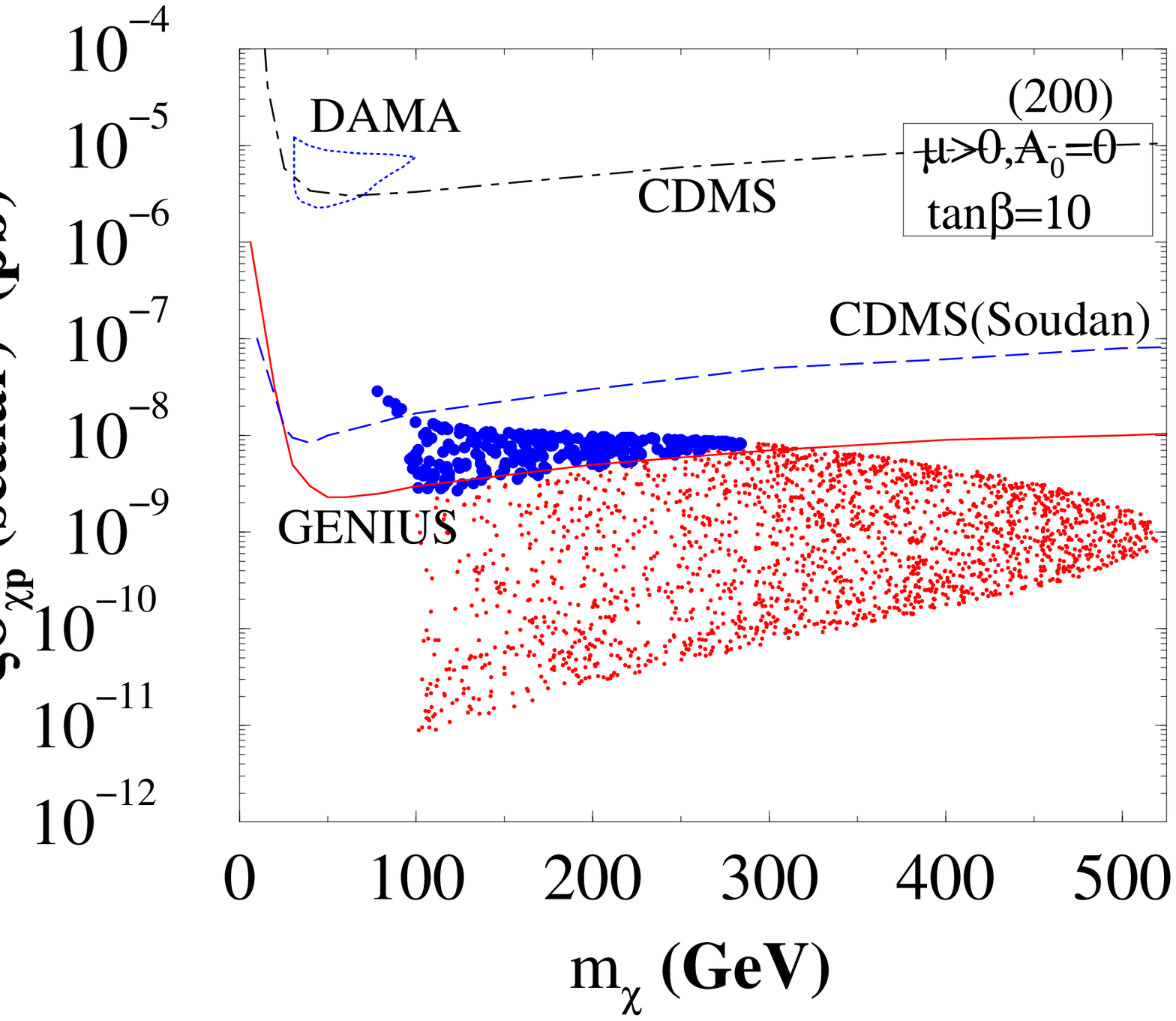}
\end{minipage}}
\hspace*{0.3in}                       
%\subfigure[Caption for picture two]{                       
\subfigure[]{                       
\label{sigma_d}                       
\begin{minipage}[b]{0.5\textwidth}                       
\centering                      
\includegraphics[width=\textwidth,height=0.8\textwidth]{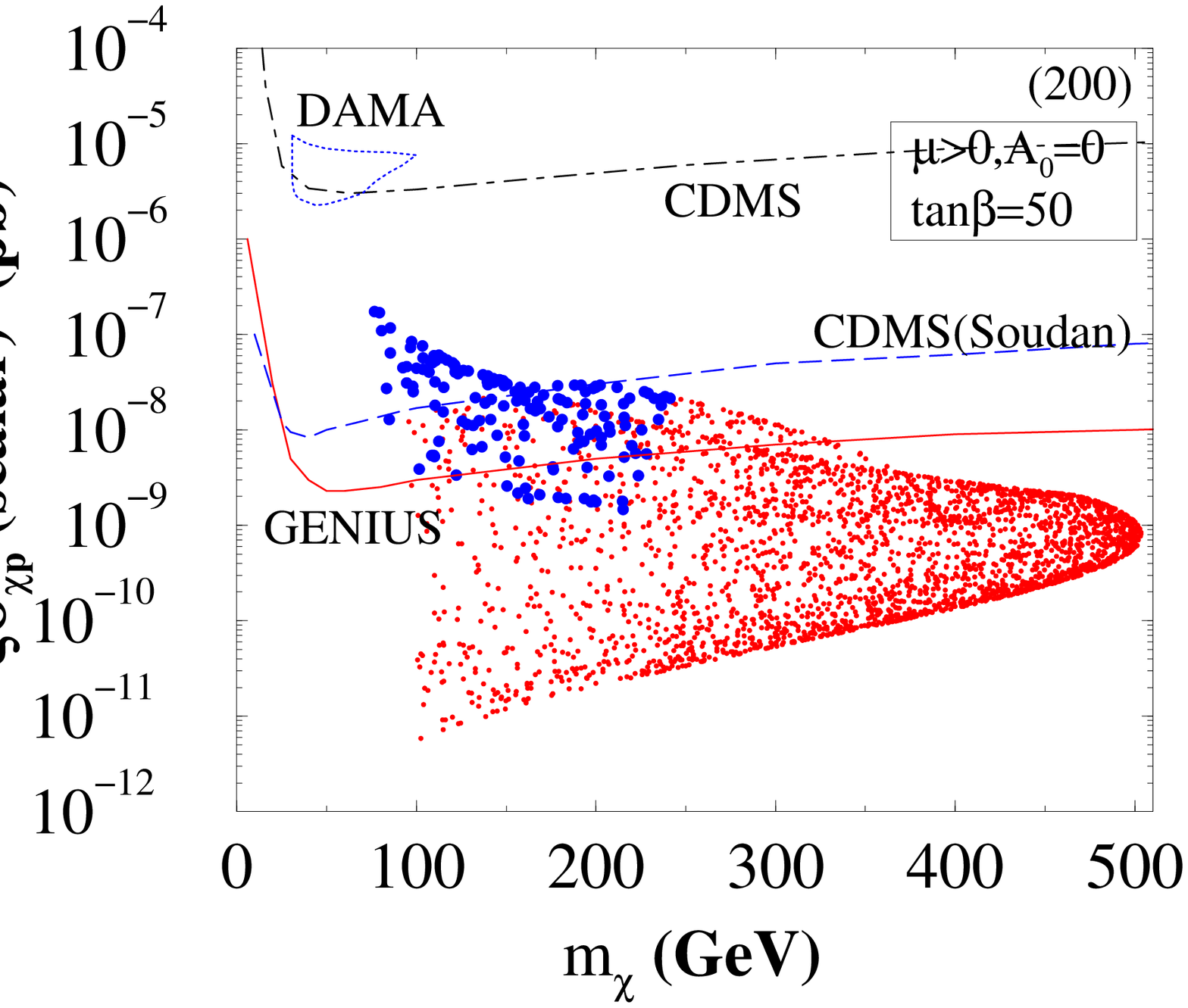}    
\end{minipage}}                       
\caption{(a) and (b): Spin independent neutralino-proton 
scattering cross sections $\sigma_{{\tilde {\chi_1^0}}-p}$  
vs $m_{{\chi_1}^0}$ for 
$\tan\beta=$ 10 and 50 respectively for the 
$n=200$ nonuniversal gaugino mass model.  The (red) dots and the (blue) filled 
circles refer to points with below and above 10\% gaugino 
fractions respectively in this generic Higgsino dominated 
LSP scenario.
(c) and (d): Same as (a) and (b) except for the scaled spin independent 
cross sections $\xi \sigma_{ {\tilde \chi}_1^0 -p }$ 
where $\xi=
{\Omega_{\tilde \chi} h^2}/
{<\Omega_{\tilde \chi} h^2>}_{min}$, with 
${<\Omega_{\tilde \chi} h^2>}_{min}=0.05$.}
\label{sigma} 

\end{figure}

The spin -independent (scalar) cross-section is dominated by 
Higgs exchanges.  Since the Higgs couplings to a ${\tilde \chi}^0 
{\tilde \chi}^0$ pair is proportional to the product of their Higgsino 
and gaugino components~\cite{Guchait:1994zk}, they are suppressed 
for both Higgsino and gaugino dominated DM.  Fig.~\ref{sigma} gives 
scatter plots of the scalar cross-section with and without rescaling for
$\tan\beta=$10 and 50.  The upper range of the scatter-plots 
correspond to the mixed Higgsino-gaugino region ($Z_g >10 \%$) of 
Fig.~\ref{relic_figs}, as expected.  The cross sections without the rescaling 
factors are moderately larger than the minimal SUGRA 
prediction~\cite{Ellis:2001pa}.  But the rescaled cross-sections are similar 
in size to the latter.  Fig.~\ref{sigma} also shows that a significant 
part of the unrescaled cross-section lies above the discovery limits 
of the future CDMS~\cite{futureCDMS} and GENIUS~\cite{genius} 
experiments; but the rescaled cross-sections 
generally lie below these limits except for the mixed Higgsino-gaugino region.  The DAMA~\cite{dama} and 
present CDMS limits are also shown.  In other words the upcoming experiments 
can detect the Higgsino DM if it is the dominant component of the 
galactic DM, but not if it is only a subdominant component of the latter.  
\section{Summary}
We have investigated the dark matter phenomenology of a SUGRA model 
with nonuniversal gaugino masses.  Its gauge kinetic function is a function 
of a nonsinglet chiral superfield, belonging to the 200-plet representation 
of SU(5).  It is as predictive as the minimal SUGRA model and has less 
fine-tuning problem than the latter.  It predicts a dominantly Higgsino 
LSP over the practically entire parameter space.  The resulting thermal 
relic density of the Higgsino dark matter lies moderately below 
the cosmologically favoured range of eq.(\ref{cdmlim}).  Thus the Higgsino 
can only be a subleading component of the cold dark matter in the 
standard cosmological scenario.  On the other hand its unsuppressed 
coupling to the Z-boson implies an enhanced rate of capture by 
the Sun.  Consequently the predicted rate of indirect detection via 
high energy neutrinos coming from its pair-annihilation inside the 
Sun is much larger than the minimal SUGRA even after rescaling by the 
low density factor.  This signal can be detected by the proposed 
${\rm km^2}$ neutrino telescopes like IceCube and ANTARES over practically 
the full parameter space of the model.  For the direct detections 
the predicted rate after rescaling is rougly similar to the minimal SUGRA 
prediction. 

 We thank Manuel Drees for discussion and for the use of his radiative 
correction code for chargino and neutralinos.

\newpage

\newpage

\newpage

\newpage
\newpage

\end{document}